\documentstyle[11pt,aaspp4,flushrt]{article}  

\def\zL{{\zeta_{\rm L}}}
\def\teff{{T_{\rm visc}}}

\def\msun{{\rm M_{\odot}}}

\def\dmc{{\dot M_{\rm cr}}}
\def\lta{\la}
\def\gta{\ga}
\begin{document}

\slugcomment{Accepted for publication in the Astrophysical
Journal (May 1997)}

\title{WHY LOW--MASS BLACK HOLE BINARIES ARE TRANSIENT}

\author{A.~R.~King, U.~Kolb and E.~Szuszkiewicz}
\affil{Astronomy Group, University of Leicester, 
Leicester LE1 7RH, U.K.\\
(ark@star.le.ac.uk, uck@star.le.ac.uk, esz@star.le.ac.uk)
} 
\authoremail{ark@star.le.ac.uk, uck@star.le.ac.uk, esz@star.le.ac.uk}


\begin{abstract}
We consider transient behavior in low--mass X--ray binaries (LMXBs). In
short--period neutron--star systems (orbital period $\lta 1$~d)
irradiation of the accretion disk 
by the central source suppresses this except at very low mass transfer
rates. Formation constraints however imply that a significant fraction
of these neutron star systems  
have nuclear--evolved main--sequence secondaries and thus mass
transfer rates low enough to be transient. But most short--period
low--mass black--hole systems will form with unevolved
main--sequence companions and have much higher mass transfer
rates. The fact that essentially all of them are nevertheless 
transient shows that irradiation is weaker, as a direct 
consequence of the fundamental black--hole property -- the lack of a hard 
stellar surface. 
\end{abstract}


\section{INTRODUCTION}

The circumstantial evidence that many soft X--ray transients (SXTs)
contain black holes is now very strong. Several of these systems
have been found to have mass functions so large that the compact star's
mass exceeds any likely value for the maximum mass of a neutron star. By 
elimination it is generally believed that this star must be a black hole.
The incidence of such black--hole candidate systems (BLMXBs)
among SXTs with known
orbital periods (8 out of 14) is much greater than any likely estimate
of their incidence among persistent LMXBs (1 out of 29). It has been
suspected for some time 
that this is a consequence of the lower mass transfer rates expected
in black--hole systems (Mukai 1994). At the rates expected for 
main--sequence companions the accretion disks
in these systems would be unstable, giving rise to the X--ray outbursts
(e.g. Cannizzo et al.\ 1982, Lin \& Taam 1984). However this would also be 
true of most neutron--star systems (NLMXBs), and yet the majority of these are
observed to be persistent. In this context van Paradijs (1996) pointed out the 
crucial importance of irradiation of the disk by the central accreting source
in determining its stability. For a central point source this leads to a much
tighter upper limit on the mass transfer rate allowing transient behavior,
which is satisfied by all observed NLMXBs.

Following this, King, Kolb \& Burderi (1996; hereafter KKB) and
King et al.\ (1997) showed that this more stringent condition still makes
most long--period NLMXBs with
a low--mass giant donor (orbital period $P \ga 1$~d) transient,
whereas short--period NLMXBs with a main--sequence donor ($P \lta 1$~d)
are transient only if the companion star is significantly
nuclear--evolved before 
mass transfer begins. The fact that a non--negligible fraction of such systems 
are nevertheless observed to be transient imposes very tight 
constraints on the formation of NLMXBs. These turn out to hold 
in practice (King \& Kolb 1997), strengthening one's belief in the 
consistency of the whole picture: NLMXBs can only form with companions
sufficiently evolved that many of them have low mass transfer rates and 
appear as SXTs. However, while irradiation brings clarity
to the picture of NLMXBs, it appears to complicate matters for 
the black--hole systems. For here the formation constraints are far weaker
than for NLMXBs. In particular survival of the binary after 
any supernova explosion
forming the compact object is virtually guaranteed for a black hole,
in stark contrast to the 
neutron--star case; there is no apparent reason why most short--period
BLMXBs should not 
form with completely unevolved companion stars. These systems would have mass
transfer rates well above the limit for transient behavior in NLMXBs. This
fact would apparently force us to
predict the existence of large numbers of persistent BLMXBs, quite contrary
to observation. 

We address this problem here. We shall show that the resolution of the apparent
paradox is that irradiation differs sharply for neutron--star and black--hole
systems. In the latter case, the lack of a hard stellar surface means that
the central source is the inner region of the accretion disk itself. The
small solid angle subtended by these regions at the outer parts of the disk
greatly weakens the stabilizing effect of irradiation, allowing 
essentially all BLMXBs to be transient. It thus appears that transient behavior
gives direct confirmation of the fundamental black--hole property precisely
in systems whose mass estimates already suggest the presence of black holes.

\section{DISK INSTABILITIES IN NLMXBs}

The conditions for disk instability in LMXBs have been considered in detail by 
van Paradijs (1996), KKB, and King et al.\ (1997), and we summarize their
conclusions here. If a steady state exists in which 
all of the disk is above the hydrogen ionization temperature $T_H\sim 6500$~K,
the instability will be suppressed and the source will appear as persistent.
Viscous dissipation alone gives an effective temperature $\teff$, with 
(e.g. Frank et al.\, 1992)
\begin{equation}
T_{\rm visc}^4 = {3GM\dot M \over 8\pi \sigma R^3} \label{eq1}
\end{equation}
at disk radii $R$ much larger than that of the central object ($M$ is
the mass of this star, $\dot M$ the accretion rate, $G$ the
gravitational constant and $\sigma$ the Stefan--Boltzmann
constant). 
Irradiation plays an important role in potentially raising the disk's
surface temperature $T$ above $\teff$, i.e.
\begin{equation}
T^4 = T^4_{\rm visc} + T_{\rm irr}^4 \, .
\label{eq2}
\end{equation}
For a point source at the center of the disk, the irradiation temperature 
$T_{\rm irr}$ is given by
\begin{equation}
T_{\rm irr}^4 = {\eta \dot Mc^2(1-\beta)\over 4\pi \sigma R^2}{H\over R}
\biggl({{\rm d}\ln H\over {\rm d}\ln R} - 1\biggr). \label{eq3}
\end{equation}
Here $\eta$ is the efficiency of rest--mass energy conversion into X--ray 
heating, $\beta$ is the X--ray albedo, and $H(R)$ is the local disk
scale height. Since $T$ always decreases with $R$, 
the condition $T > T_H$ for steady accretion is
most stringent at the outer edge $R_d$ of the disk. We define a critical 
accretion rate $\dmc$ by the equation
\begin{equation}
T(R_d) = T_H \label{eq4}.
\end{equation}
Formally we define also the critical rate $\dmc^{\rm visc}$ in the
absence of irradiation ($\teff(R_d) = T_H$), and the critical rate 
$\dmc^{\rm irr}$ for fully irradiation--dominated disks
($T_{\rm irr}(R_d) = T_H$). These definitions imply that 
\begin{equation}
 \frac{1}{\dmc} = \frac{1}{\dmc^{\rm irr}} + \frac{1}{\dmc^{\rm visc}}
 \, .
\label{eq4a}
\end{equation}

Thus the disk instability is suppressed for $\dot M > \dmc$ and
we can try to discriminate between steady and outbursting 
systems by checking this condition, equating the accretion rate $\dot M$
to the mass transfer rate $-\dot M_2$ from the companion. 
In CVs, $T_{\rm irr} \ll \teff$, and
this procedure correctly predicts that all disk--accreting systems below
the period gap should be dwarf novae. In NLMXBs $T_{\rm irr}$ is 
all--important: the last two factors on the rhs of (\ref{eq3})
are typically $\propto H/R \sim $ constant (see below), so 
$T_{\rm irr}^4$ falls off only as $R^{-2}$ compared with $R^{-3}$ for 
$T_{\rm eff}^4$. At the outer disk edge $R_d$ we thus expect $T_{\rm irr} >
\teff$, 
giving a critical accretion rate $\dmc$ which is typically almost two
orders of magnitude lower at a given orbital period for main--sequence
donor systems (see below and Fig.~3). 

It is sometimes suggested that the outer disk is not affected by
irradiation, either because the disk is convex and shadows its outer
parts, or because the outer layers form an optically thin corona
above the disk (e.g.\ Tuchman, Mineshige \& Wheeler 1990; Cannizzo
1994). In both cases $T \simeq T_{\rm visc}$ rather than (\ref{eq2}).
However, the fact that $\dmc$ (from
eq.~\ref{eq4a}) rather than $\dmc^{\rm visc}$ separates transient 
from persistent NLMXBs (van Paradijs 1996), as well as the high
optical to X--ray flux ratio observed in LMXBs, is strong evidence
for the dominant role of irradiation. 

Thus assuming that the outer disk temperature is controlled by
irradiation, we conclude that short--period neutron--star
transients must have relatively 
low mass transfer rates. KKB deduced that they must have companions which
are nuclear--evolved before mass transfer begins, even at very short 
periods where one might otherwise expect a completely unevolved
lower--main--sequence companion. 
Since a significant fraction of NLMXBs are transient this peculiar condition
must follow from the constraints on the formation of NLMXBs. 
King \&  Kolb (1997) indeed show that this is true if the average kick
velocity imparted on the neutron star at birth is small. 
To prevent disruption the binary has to retain at least one--half of
its mass after the helium star supernova (SN) which forms the neutron
star. The neutron star probably has a mass close to $1.4 \msun$
at birth, favoring the survival of systems with a massive
companion and a low mass neutron star progenitor. 
To accommodate such a low mass helium star, the pre--SN orbit must have
been very wide, hence also the post--SN orbit. The corresponding 
long detached post-SN evolution causes the secondary 
to be nuclear--evolved at turn--on of mass transfer.

By contrast it is clear that no such 
constraint can hold for black--hole systems: even if a supernova occurs,
which is unclear, the black hole can easily bind more than one--half of the
progenitor mass, thus leaving  the companion mass effectively
unconstrained. Hence we expect that
most short--period BLMXBs will form with unevolved main--sequence
companions. Magnetic braking and gravitational radiation will then
drive mass transfer rates considerably 
higher than the $\dmc$ deduced for NLMXBs above, and one would
at first sight expect the vast majority of short--period BLMXBs
to be persistent X--ray sources, in complete contradiction to what is
observed.

\section{DISK INSTABILITIES IN BLMXBs}

The answer to this difficulty
is the fact that irradiation is much weaker if the accreting object is a 
black hole rather than a neutron star. This is already mentioned in
the seminal paper on accretion disk structure by Shakura \& Sunyaev (1973).
The reason is that in the black--hole case the central object has no hard
surface, and so cannot act as a point source for irradiation as assumed in the
derivation of (\ref{eq3}). Of course, a comparable luminosity is released in
the inner part of the accretion disk surrounding the hole, but this is a 
flat surface lying in the disk's central plane, and thus almost parallel to
the surface layers of the outer regions of the disk. A full calculation is
complex, but it is clear that the solid angle subtended by the inner disk is
smaller by a factor $\sim H/R$ than assumed in (\ref{eq3}). 
We can neglect relativistic beaming effects for irradiation of the
outer disk (Cunningham 1976, Zhang et al.\ 1997).
For simplicity we therefore replace (\ref{eq3}) by
\begin{equation}
T_{\rm irr}^4 = {\eta \dot Mc^2(1-\beta)\over 4\pi \sigma R^2}
\biggl({H\over R}\biggr)^2
\biggl({{\rm d}\ln H\over {\rm d}\ln R} - 1\biggr) \label{eq5}
\end{equation}
in the black--hole case. With this change, (\ref{eq4}) now defines a new
(higher) critical mass transfer rate $\dmc$, which we must
compare with the rates expected from binary evolution. 

To evaluate $\dmc^{\rm irr}$ we assume that $R_d$
is about 70\% of the primary's Roche lobe radius $R_L$, which in turn
is a fraction $f_1$ of the binary separation $a$; $f_1$ depends only
on the ratio $q = M_2/M_1$ of donor mass $M_2$ to primary mass $M_1$. 
For $T_{\rm irr}$ we need an estimate of $H(R)$. Neglecting
internal viscous dissipation and assuming an isothermal vertical disk
structure it is easy to show (e.g.\ Cunningham 1976, Fukue 1992) that
$H \propto R^{9/7}$ if $T_{\rm irr}$ is given by (\ref{eq3}), 
and $H \propto R^{4/3}$ with $T_{\rm irr}$ from (\ref{eq5}).  
In reality the disk is not isothermal and the 
slope is closer to the standard law $H\propto R^{9/8}$ for
unirradiated discs. We adopt $43/36$ and $45/38$ for the BH case
(\ref{eq5}) and the neutron star case (\ref{eq3}), respectively
(cf.\ Burderi, King \& Szuszkiewicz 1997). 
Furthermore, 
observations of LMXBs indicate $H/R \simeq 0.2$ and $\beta\simeq 0.9$
(de Jong et al, 1996, and references therein). 
Using these results and $\eta=0.2$ in (\ref{eq4}, \ref{eq5}) gives
\begin{equation}
\dmc^{\rm irr} = 
1.34 \times 10^{-10} \: f_1^2 (m_1 + m_2)^{2/3} \: P_h^{4/3} \:
\msun {\rm yr}^{-1} \, ,
\label{eq6}
\end{equation}
with $m_1 = M_1/\msun$, $m_2=M_2/\msun$ and $P_h=P/{\rm hr}$.

\section{MASS TRANSFER RATES IN BLMXBs}

At periods longer than about a day, mass transfer in LMXBs
is driven by the nuclear expansion of the
secondary. For low--mass systems with this star on the first giant branch
a simple core--envelope description is available, and gives the mass transfer 
rate as
\begin{equation}
- \dot M_2 = 
\frac{7.26 \times 10^{-10}}{\zeta_{\rm eq} - \zL} \:
m_2^{1.46} \: P_d^{0.93} \: \msun {\rm yr}^{-1}
\label{eq9}
\end{equation}
(cf.\ King et al.\ 1997, King 1988).
Here $P_d$ is the orbital period in days, $\zeta_{\rm eq}$ the thermal
equilibrium mass--radius exponent (taken as $0$) and $\zL \simeq
2M_2/M_1 - 5/3$ the secondary's Roche lobe index.
Transient behavior requires $- \dot M_2 < \dmc$, which
from (\ref{eq1}), (\ref{eq4a}), (\ref{eq6}) and (\ref{eq9}) translates
into a lower limit on 
$M_1/M_2$ for given $P$ and $M_2$ (Fig.~1). The only persistent BH systems  
are those with an unusually low BH mass ($\lta 3 \msun$) and high  
companion mass ($\gta 1.5 \msun$), putting them
close to mass transfer instability. Such systems evolve quickly to
the transient regime as the companion mass is reduced. This 
can be seen in Fig.~2, where we show 
the mass transfer rate as a function of orbital period along typical 
evolutionary sequences. This of course is not surprising in the light
of the results of King et al.\ (1997) who 
showed that the NLXMB prescription (\ref{eq3}) already makes most 
systems transient. 
We conclude that essentially {\it all long--period BLMXBs should be 
transient}.

\medskip

At short binary periods mass transfer is driven by angular momentum 
loss via magnetic braking and gravitational radiation. 
In contrast to the nuclear--evolution case the mass transfer rate
here is quite sensitive to the primary mass $M_1$. 
First,
larger $M_1$ raises the orbital angular momentum of the binary as
$J \propto M_1^{2/3}$ for $M_1 \gg M_2$: 
if the angular momentum loss rate $\dot J$
is independent of $M_1$ (as is the case for magnetic stellar wind 
braking) the transfer rate is reduced as $-\dot M_2 \propto M_1^{-2/3}$, 
On the other hand we know that angular momentum loss
via gravitational radiation actually increases, with 
$-\dot M_2 \propto M_1^{2/3}$.
Using the form of Verbunt \& Zwaan (1981) for the magnetic braking rate,
with the radius of
gyration set to $(0.2)^{1/2}$ and the calibration parameter to unity,
and the standard form (e.g. Landau \& Lifschitz 1958) for the gravitational
radiation losses, the mass transfer rate is 
$-\dot M_2 = \dot M_{\rm MB} + \dot M_{\rm GR}$, with 
\begin{equation}
   \dot M_{\rm MB} = 
  \frac{1.08 \times 10^{-7}}{\zeta_{\rm eq} - \zL} \: \frac{(m_1 + m_2)^{1/3}
  m_2^{7/3}}{m_1} \: P_h^{-2/3} \: \msun {\rm yr}^{-1} \, ,
\label{eq10a}
\end{equation}
and 
\begin{equation}
   \dot M_{\rm GR} = \frac{2.52 \times 10^{-8}}{\zeta_{\rm eq} - \zL} \:
  \frac{m_1 m_2^2}{(m_1 + m_2)^{1/3}} \: P_h^{-8/3} \: 
  \msun {\rm yr}^{-1}   \, .
    \label{eq10b}
\end{equation}
($\zeta_{\rm eq} \simeq 1$, $\zL \simeq 2M_2/M_1 - 5/3$ as above).

In Fig.~3 we plot the mass transfer rate and critical rate
(\ref{eq6}) as a function of orbital period for representative
evolutionary sequences with an unevolved secondary for 
different initial BH mass $m_1 = 2, 5$ and $10$. 
Magnetic braking is assumed to operate only as long as the secondary
has a radiative 
core. This leads to a detached phase (``period gap'') when the
secondary becomes fully convective, followed by the resumption of mass
transfer below the period gap. 
BLMXBs with unevolved donors above the detached phase ($P \ga 3$~h) 
are transient if the BH mass is $\gta 5\msun$, whereas at shorter periods  
($P \lta 2$~h) BLMXBs are transient for any reasonable BH mass. 
We note, however, that the critical transfer rate is only slightly 
higher than the secular mean transfer rate in systems close to the
period minimum and the upper edge of the detached phase for all BH
masses. In view of the uncertainties in (\ref{eq6}), one would
therefore not be surprised by the appearance of persistent BLMXBs 
(even with a $10\msun$ primary) at these periods.

We can get a rough analytic representation of this result by introducing a 
number of simplifications:
Using 
\begin{equation}
  f_1 \simeq f_2 {\left( \frac{M_1}{M_2} \right)}^{5/12} 
\label{eq7}
\end{equation}
the ratio $f_1$ can be obtained to better than $8\%$ for $0.001 <
M_2/M_1 < 1$ from $f_2$, the secondary's Roche lobe radius in units of
$a$, which in turn is given by 
$f_2 \simeq (8 M_2/81(M_1+M_2))^{1/3}$ (Paczy\'nski 1971). 
Thus we have from (\ref{eq6})
\begin{equation}
\dmc^{\rm irr} \simeq
2.86 \times 10^{-11} \: m_1^{5/6} \: m_2^{-1/6} \: P_h^{4/3} \:
\msun {\rm yr}^{-1} \, .
\label{eq7a}
\end{equation}
We then set $\dmc = \dmc^{\rm irr}$, $\zeta_{\rm eq} - \zL =8/3$ and
assume $m_1 \gg m_2$.

For systems above the gap ($P_h\ga 3$) we furthermore assume 
$-\dot M_2 \simeq \dot M_{\rm MB}$ and define
the parameter $\hat{m_2} = M_2/M_2({\rm MS})$, where $M_2({\rm MS})$
denotes the mass of a main--sequence secondary which fills its Roche
lobe at a given orbital period ($P_h \simeq 8.1 M_2({\rm MS})/\msun$). 
Then we find from (\ref{eq7a}) and (\ref{eq10a}) as a condition for
transient behavior  
\begin{equation}
 m_1 > 3.9 \: \hat{m_2}^{5/3} \: P_h^{1/3} \, .
\label{eq11}
\end{equation}
Thus the minimum primary mass becomes even smaller if the
main--sequence secondary is somewhat nuclear--evolved ($\hat m_2 
< 1$). 

For systems with a fully convective secondary ($P_h \lta 2$) we have 
$-\dot M_2 = \dot M_{\rm GR}$. Detailed models show that along the
pre--period minimum evolution the relation $P_h \simeq  k \:
m_2^{2/3}$ holds, for 
any BH mass. The constant $k$ is $\simeq 5.8$ for secondaries with a
helium content $Y=0.28$, and slightly larger for larger $Y$ (as 
would be the case if the secondary was nuclear--evolved
at turn--on of mass transfer at longer $P$). Hence 
we find from (\ref{eq7a}) and (\ref{eq10b}) 
\begin{equation}
P_h \: m_1^{2/9} > 1.1 \: {\left( \frac{5.8}{k} \right)}^{13/3}
\label{eq12}
\end{equation}
as a condition for transient behavior, insensitive to $M_1$ and always
fulfilled.  

\medskip

It has recently been suggested that black hole SXTs might be inefficient
accretors in quiescence. The inner part of the accretion disk would be
replaced by a
quasi--spherical (``advective'') accretion flow taking the matter
into the black hole before it has time to radiate (Narayan,
McClintock \& Yi 1996; cf.\ Katz 1977; Begelman 1978). It is unclear how
the SXT outbursts themselves would be produced in this picture; but 
nevertheless it is important to note that our criterion $\dot M <
\dot M_{\rm cr}$ refers only to the {\it mean} accretion rate, i.e. a
system is assumed to be transient if steady disk flow with an accretion
rate equal to the mass transfer rate is unstable. Our considerations here
would only be affected if this (hypothetical) steady flow was
itself advection--dominated. In this case there would clearly be
no question of irradiation stabilizing the disc. Thus our
conclusion that black hole disks are unstable because the
fundamental property of these objects weakens the irradiation effect
would be still more strongly justified.

\section{CONCLUSIONS}

Irradiation of the accretion disk surface by the central source
has a determining effect on transient behavior. In neutron--star
low--mass X--ray binaries this confines transient behavior to systems 
with extremely low accretion rates. The very strong formation 
constraints for neutron--star binaries with low--mass companions
nevertheless mean that a significant fraction of these systems are
transient. The much weaker contraints on forming black--hole systems
would at first sight suggest that very few of them would have mass
transfer rates low enough to be transient. The fact that the vast
majority of black--hole candidates are nevertheless transient shows
that irradiation is weakened by the fundamental property
that black holes have no hard surface; that is, they are black.

\acknowledgements

Theoretical astrophysics research at Leicester is supported by a PPARC
Rolling Grant. ARK gratefully acknowledges the award of a PPARC Senior
Fellowship.

\clearpage
 
\section*{FIGURE CAPTIONS}
\bigskip

\figcaption{
Minimum ratio $M_1/M_2$ (hence lower limit on the BH mass $M_1$) 
for SXT behavior in BLMXBs with giant donors as a function of orbital
period $P$ (in days), for different donor mass (upper solid line:
$M_2=2\msun$, lower solid line: $M_2=1\msun$), assuming $\zeta_{\rm
eq}=0$. 
Mass transfer stability ($\zeta_{eq} > \zL$) requires that $M_1/M_2$
is above the dashed line. Systems evolve
towards longer orbital period and larger $M_1/M_2$. The evolutionary
track for the sequence shown in the lower left panel of Fig.~2 
is indicated by a dotted line.
}

\bigskip

\figcaption{
Mass transfer rate $\dot m = -\dot M_2/(\msun {\rm yr}^{-1})$ 
(solid curves) and the corresponding critical mass transfer rate
$\dmc$ (dashed) as a function of orbital period $P$ (in days) along
various evolutionary sequences of long--period BLMXBs with giant
donors. Systems are transient if the track is below the dotted
line.
The initial BH mass is $10\msun$ in the upper panel, $2\msun$ in the
lower panel. The initial core mass is $M_c=0.20\msun$ and 
$0.37\msun$ for the sequences on the left and right,
respectively. Mass transfer begins with a $1.5\msun$ secondary and 
terminates when $M_2=M_c$ (constant total mass and $\zeta_{\rm eq}=0$ 
was assumed). 
}

\bigskip

\figcaption{
Mass transfer rate $\dot m = -\dot M_2/\msun {\rm yr}^{-1}$ versus
orbital period $P$ (in h) for short--period
BLMXBs with an unevolved secondary of initial mass $1\msun$ for 
different initial primary masses ($2\msun$, $5\msun$, $10\msun$, 
from top to bottom). The total binary mass is constant along the
sequence. Magnetic braking is assumed to operate only as long as the
secondary has a radiative core. 
The smaller contrast between the angular momentum losses above and
below the period gap for increasing BH mass means that the 
detached phase at $P\simeq 3$~h (``period gap'') is narrower, and also
that the minimum period is longer. The dashed curve represents the
critical mass transfer rate for BH accretors. Systems with lower mass
transfer rates are transient.  Also shown is the critical mass
transfer rate for neutron star 
accretors (dotted; assuming $\eta=0.15$) and unirradiated disks
(dash-dotted). 
}

\newpage

\begin{figure}
\plotone{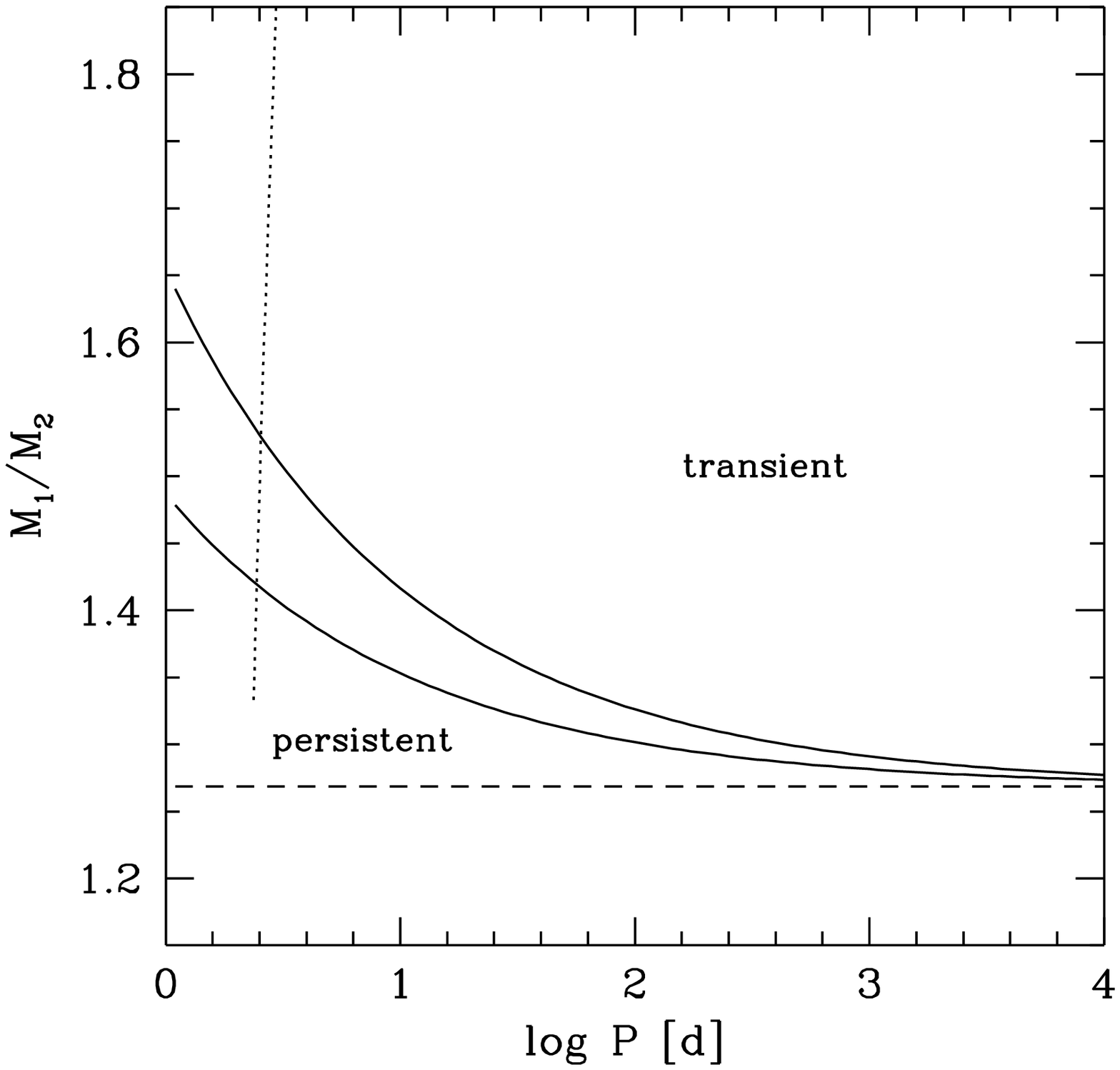}

\vspace{1.5cm}

{\Large Figure 1}
\end{figure}

\newpage

\begin{figure}
\plotone{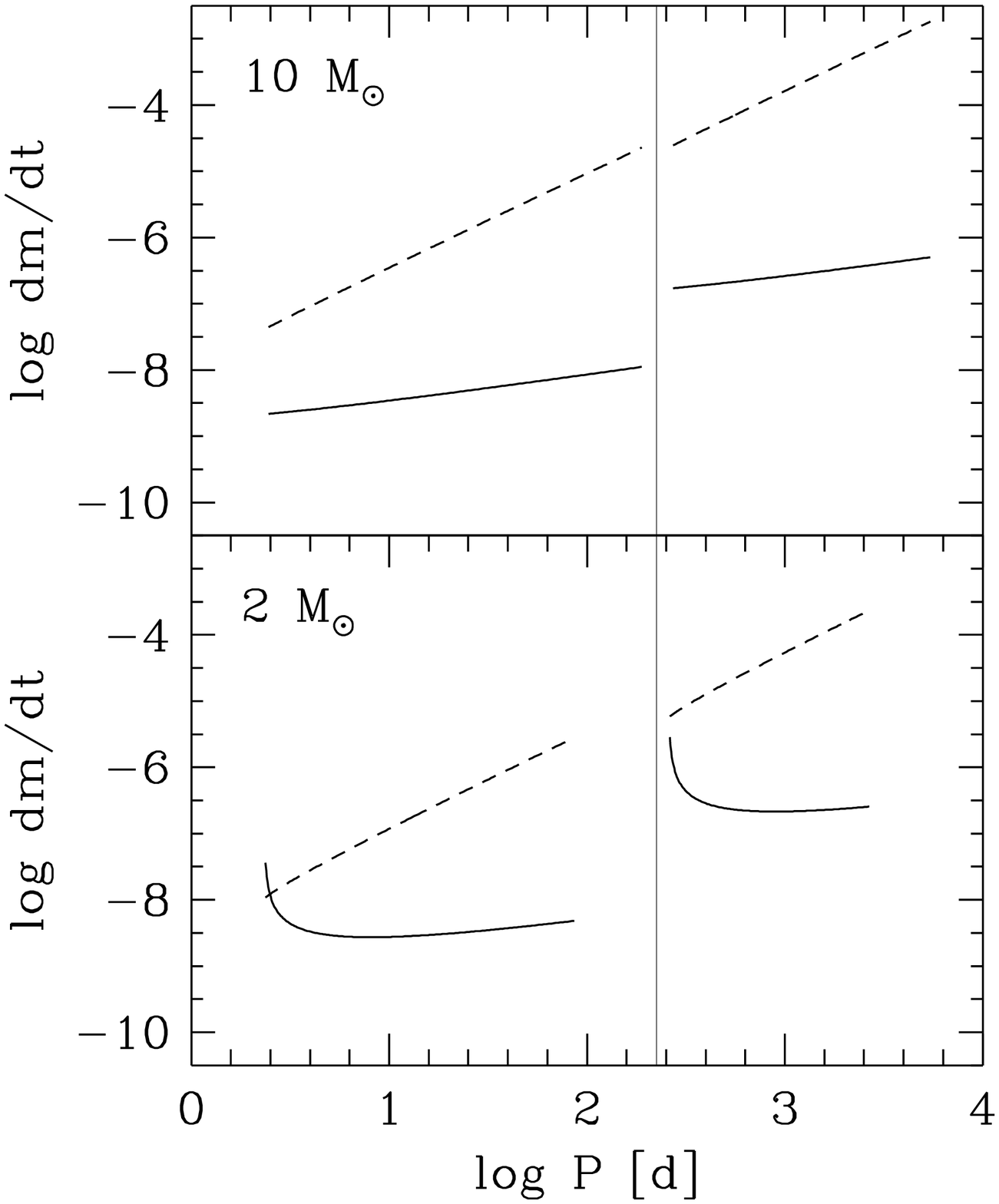}

\vspace{1.5cm}

{\Large Figure 2}
\end{figure}

\newpage

\begin{figure}
\plotone{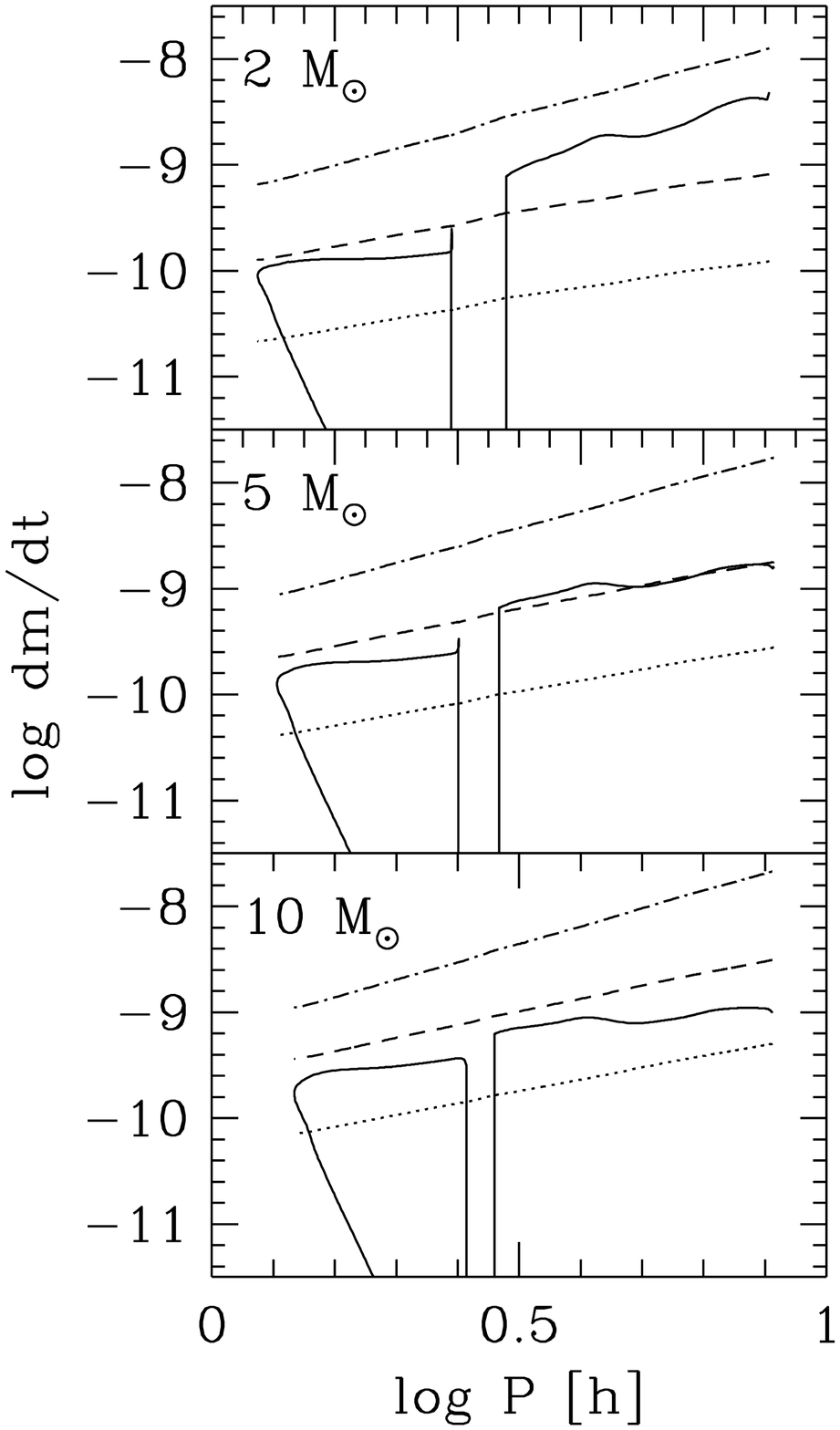}

\vspace{1.5cm}

{\Large Figure 3}
\end{figure}

\end{document}